\documentclass[final,longtitle,5p,times,twocolumn]{elsarticle}

\usepackage{graphicx}
\usepackage{dcolumn}
\usepackage{xcolor}
\usepackage{hyperref}
\usepackage{bm}
\usepackage[]{units}
\usepackage{soul} 
\usepackage{amssymb}
\usepackage{amsmath}
\usepackage{comment}
\usepackage[T1]{fontenc} 
\usepackage{array}
\usepackage[left]{lineno}
\usepackage{mwe}
\usepackage{listings}
\usepackage{ulem} 
\usepackage{fontawesome5}

\definecolor{mygreen}{rgb}{0,0.6,0}
\definecolor{mygray}{rgb}{0.5,0.5,0.5}
\definecolor{mylightgray}{rgb}{0.937,0.937,0.937}
\definecolor{mymauve}{rgb}{0.58,0,0.82}

\newcommand*\boldell{\pmb{\ell}}

\journal{Computer Physics Communications}

\begin{document}
\begin{frontmatter}

\title{{\sc GRANDlib}:\\A simulation pipeline for the Giant Radio Array for Neutrino Detection (GRAND)~\href{https://github.com/grand-mother/grand}{\faGithubSquare}\\ \vspace{0.25cm}\normalsize
(The GRAND Collaboration\tnoteref{email})
\vspace{-0.4cm}}
\tnotetext[email]{collaboration@grand-observatory.org}

\author[1,2]{R. Alves Batista}
\author[3]{A. Benoit-Lévy}
\author[4,5]{T. Bister}
\author[6]{M. Bohacova}
\author[7]{M. Bustamante}
\author[8]{W. Carvalho}
\author[9,10]{Y. Chen}
\author[11]{L. Cheng}
\author[12]{S. Chiche}
\author[2]{J. M. Colley}
% ...

\address[1]{Institut d'Astrophysique de Paris, CNRS  UMR 7095, Sorbonne Université, 98 bis bd Arago 75014, Paris, France}
\address[2]{Sorbonne Université, Université Paris Diderot, Sorbonne Paris Cité, CNRS, Laboratoire de Physique 5 Nucléaire et de Hautes Energies (LPNHE), 6 4 place Jussieu, F-75252, Paris Cedex 5, France}
\address[3]{Université Paris-Saclay, CEA, List, F-91120 Palaiseau, France}
\address[4]{Institute for Mathematics, Astrophysics and Particle Physics, Radboud Universiteit, Nijmegen, the Netherlands}
\address[5]{Nikhef, National Institute for Subatomic Physics, Amsterdam, the Netherlands}
\address[6]{Institute of Physics of the Czech Academy of Sciences, Na Slovance 1999/2, 182 00 Prague 8}
\address[7]{Niels Bohr International Academy, Niels Bohr Institute, University of Copenhagen, 2100 Copenhagen, Denmark}
\address[8]{Faculty of Physics, University of Warsaw, Pasteura 5, 02-093 Warsaw, Poland}
\address[9]{Key Laboratory of Dark Matter and Space Astronomy, Purple Mountain Observatory, Chinese Academy of Sciences, 210023 Nanjing, Jiangsu, China}
\address[10]{School of Astronomy and Space Science, University of Science and Technology of China, 230026 Hefei Anhui, China}
\address[11]{National Astronomical Observatories, Chinese Academy of Sciences, Beijing 100101, China}
\address[12]{Inter-University Institute For High Energies (IIHE), Université libre de Bruxelles (ULB), Boulevard du Triomphe 2, 1050 Brussels, Belgium}
% ...

\newpage

\begin{abstract}

The operation of upcoming ultra-high-energy cosmic-ray, gamma-ray, and neutrino radio-detection experiments, like the Giant Radio Array for Neutrino Detection (GRAND), poses significant computational challenges involving the production of numerous simulations of particle showers and their detection, and a high data throughput.  {\sc GRANDlib} is an open-source software tool designed to meet these challenges. Its primary goal is to perform end-to-end simulations of the detector operation, from the interaction of ultra-high-energy particles, through---by interfacing with external air-shower simulations---the ensuing particle shower development and its radio emission, 
to its detection by antenna arrays and its processing by data-acquisition systems. Additionally, {\sc GRANDlib} manages the visualization, storage, and retrieval of experimental and simulated data. We present an overview of {\sc GRANDlib} to serve as the basis of future GRAND analyses.

\end{abstract}


\end{frontmatter}



\section{Introduction}
\label{sec:intro}

The Giant Radio Array for Neutrino Detection (GRAND)~\cite{GRAND:2018iaj, deMelloNeto:2023zvk, Kotera:2024iyk} is a planned  detector of ultra-high-energy (UHE) cosmic particles, with energies exceeding $10^{17}$~eV.  Its primary goal is to find the long-sought origin of UHE cosmic rays (UHECRs), purportedly the most energetic cosmic accelerators~\cite{AlvesBatista:2019tlv}.  GRAND will achieve this directly by detecting large numbers of UHECRs and, indirectly, by being sensitive to the potentially tiny fluxes of associated ultra-high-energy (UHE) gamma rays and, especially,  neutrinos, first predicted in 1969~\cite{Berezinsky:1969erk, Ackermann:2022rqc}.

GRAND uses sparse, vast ground arrays of antennas to detect the radio emission from extensive air showers triggered by UHE particles~\cite{Huege:2016veh, Schroder:2016hrv} that interact in the atmosphere or just below the surface, including in surrounding mountains.  Because radio waves have a long attenuation length in air, the arrays monitor large volumes of air and ground, conferring GRAND the large exposure needed to detect UHE particles even if their fluxes are low, and the sub-degree angular resolution needed to pinpoint sources of UHE neutrinos~\cite{Ackermann:2022rqc, MammenAbraham:2022xoc, Guepin:2022qpl}.
When looking for UHE neutrinos, GRAND targets air showers initiated by Earth-skimming tau neutrinos ($\nu_\tau$), which, on average, survive traveling underground more often than electron and muon neutrinos~\cite{Fargion:1999se}.

GRAND has a staged construction plan, each stage involving progressively larger arrays~\cite{GRAND:2018iaj, deMelloNeto:2023zvk, Kotera:2024iyk}.  An array is made up of detection units (DUs), each consisting of a passive antenna that detects the radio emission from air showers and electronics to amplify, filter, and digitize it. In its final stage, GRAND is envisioned as a collection of geographically separate arrays containing thousands of DUs each~\cite{GRAND:2018iaj, deMelloNeto:2023zvk}.

Presently, three prototype GRAND arrays are in operation
~\cite{Ma:2023siw, deMelloNeto:2023zvk, Correa:2023maq, Kotera:2024iyk, Chiche:2024ohe}: GRANDProto300 in Dunhuang, China, with its first 48 DUs out of 300 deployed, to be expanded into a 200-km$^2$ array in the near future; GRAND@Auger at the site of the Pierre Auger Observatory in Malarg\"ue, Argentina, with 10 DUs; and GRAND@Nan\c cay, at the Nan\c cay Radio Observatory, in France, with 4 DUs.
Their goal is twofold: to test the performance and deployment of the GRAND DUs under field conditions and to validate the autonomous detection of extensive air showers---i.e., large particle showers that develop in the atmosphere---reaching the array from near-horizontal directions, as expected from UHE neutrinos.

Two obstacles complicate the detection of radio emission from UHE particles: first, the presence of man-made and natural radio sources and, second, the limitations of the detector instrumentation.  Thus, to confidently claim the detection of showers made by UHE particles and to characterize them, GRAND relies on extensive computer simulations that include realistic backgrounds and detector features.

In addition, GRAND is expected to have a high throughput of data.  We estimate that GRANDProto300 alone will generate 1.1 petabytes (PB) of data during its construction, of which 400 terabytes (TB) will be simulations, \unit[450]{TB} will be collected by the array, and \unit[250]{TB} will be produced by post-processing analyses. Once GRANDProto300 is completed, we estimate the data volume to grow to \unit[2.5]{PB} per year.  [Out of this, \unit[2]{PB} comes from an estimated event rate of \unit[100]{Hz}, 20 antennas triggered per event (in an event, the radio signal fulfills pre-defined trigger conditions), and \unit[33]{kB} of data created per antenna hit. There is an additional \unit[0.5]{PB} coming from GRAND simulations and analyses.]
Extracting physical insight from such a large volume of data requires software that is able to analyze experimental data and simulate the generation, propagation, and detection of the radio signals from extensive air showers.

{\sc GRANDlib} is an open-source software tool developed by the GRAND Collaboration to address the above needs.  In its present form, it performs end-to-end simulations of the interaction of UHE particles with matter in-air or underground, the development of the ensuing extensive air shower and the emission of its radio signal---by reading the output generated by air-shower simulators---and its detection in a GRAND array.  To do this, {\sc GRANDlib} interfaces with external, well-tested software packages {\sc DANTON}~\cite{Niess:2018opy}, {\sc TURTLE}~\cite{Niess:2019hdn}, {\sc ZHAireS}~\cite{Alvarez-Muniz:2010hbb}, and {\sc CoREAS}~\cite{Huege:2013vt}.  Complementary to this, there is ongoing work---not contained in this paper---on using {\sc GRANDlib} to infer the properties of an air shower from its detection; Refs.~\cite{Mitra:2023pha, Duan:2023rtd} contain preliminary results.

\begin{figure*}[t!]
 \centering
 \includegraphics[trim={0 11cm 0 0}, width=\textwidth]{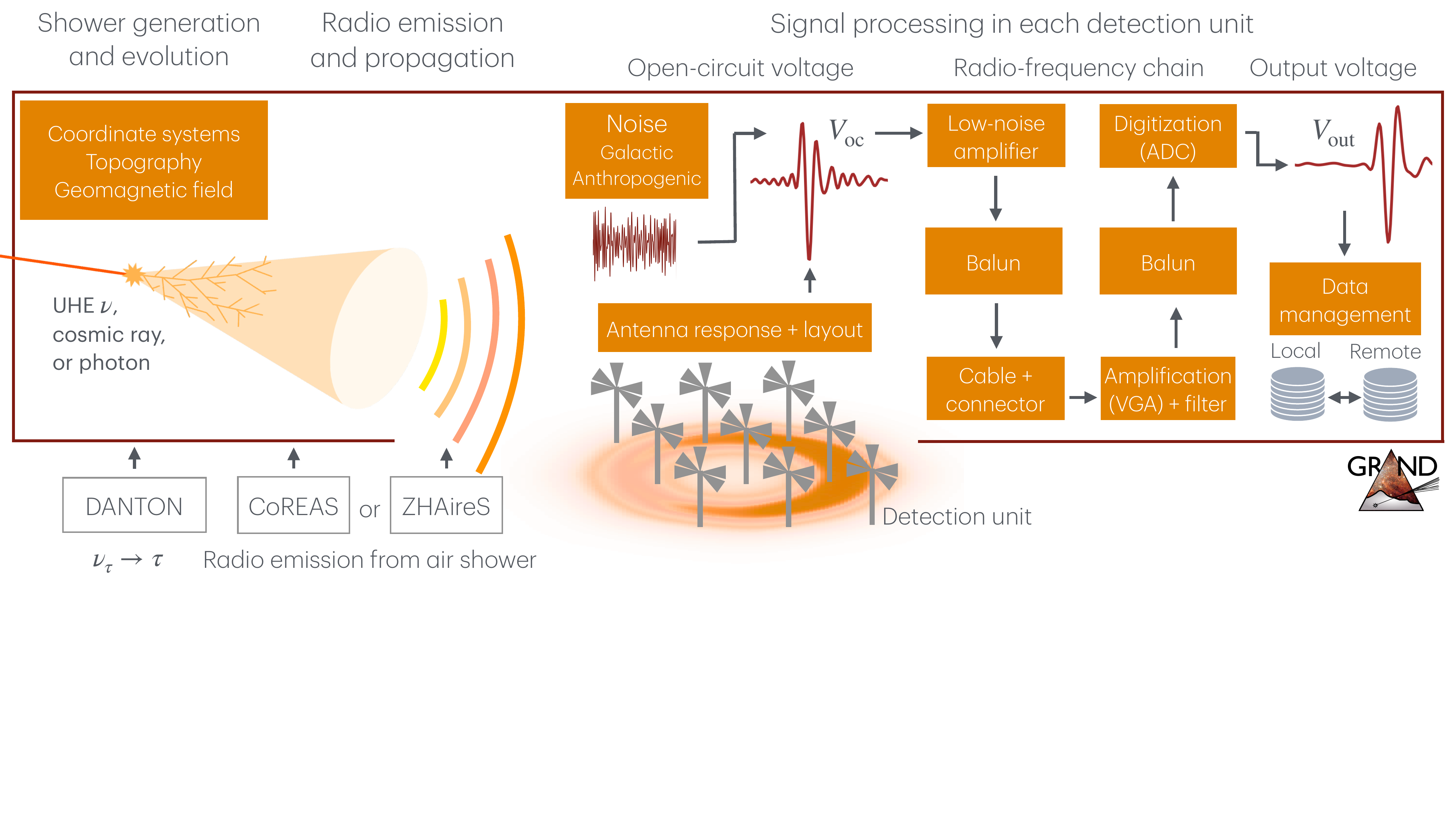}
 \caption{\textbf{Overview of the major components of {\sc GRANDlib}.} The simulation pipeline of {\sc GRANDlib} mimics closely that of the real hardware and software capabilities of a radio-detection array like GRAND~\cite{GRAND:2018iaj}.  An ultra-high-energy (UHE) particle triggers an extensive air shower whose radio emission~\cite{Huege:2016veh, Schroder:2016hrv} is picked up by an array of detection units at ground level.  In a detection unit, an antenna receives the incident radio signal and converts it into an open-circuit voltage, to which {\sc GRANDlib} adds noise.  The resulting voltage signal is then processed through the radio-frequency chain of the unit for amplification---yielding the final output voltage---digitization, and storage.  Beside the simulation pipeline, {\sc GRANDlib} has tools to handle coordinate systems, geomagnetic field, the topography of Earth, and data management.  {\sc GRANDlib} can be installed from Ref.~\cite{grandlib_repository}, which also contains usage examples.
 (The GRAND hardware includes an impedance-matching network between the antenna and the low-noise amplifier, not pictured in the radio-frequency chain presented here, but present in an imminent {\sc GRANDlib} release.)
 }
 \label{fig:grandlib_overview}
\end{figure*}

In this paper, we introduce {\sc GRANDlib}, overview its salient features, and show usage examples.  {\sc GRANDlib} is written in Python~3 and is publicly available.  Installation instructions and implementation and usage details are available in Ref.~\cite{grandlib_repository}.  All the necessary data to use {\sc GRANDlib}, including topography, geomagnetic fields, and antenna response functions are included in the {\sc GRANDlib} installation.  Although {\sc GRANDlib} has been tailored for use in GRAND, its modular design allows other experiments to adapt parts of it to suit their needs. 

The rest of the paper is organized as follows.
Section~\ref{sec:overview} gives an overview of {\sc GRANDlib}. Section~\ref{sec:coordinate} introduces its coordinate systems.  Sections~\ref{sec:topography} and \ref{sec:geomagnet} explain how {\sc GRANDlib} accesses the ground elevation and geomagnetic field at different locations.  Section~\ref{sec:shower} sketches the external air-shower simulation packages used by {\sc GRANDlib}. 
Section~\ref{sec:ant_response} explains the response of GRAND antennas.  Section~\ref{sec:signal_processing} describes signal processing, from electric fields to voltages. Section~\ref{sec:datamanagement} discusses data management. Section~\ref{sec:visualization} illustrates the data visualization capabilities of {\sc GRANDlib}. Section~\ref{sec:summary} summarizes {\sc GRANDlib} and outlines future developments.  \ref{appnd1} describes coordinate transformations between coordinate systems. \ref{appnd:fileformat} outlines the structure of the files used for simulation and experimental data. \ref{appnd4} lists the tools used to maintain code quality.  \ref{appnd5} lists external package dependencies.


\section{Overview of {\sc GRANDlib}}
\label{sec:overview}

{\sc GRANDlib} is designed to generate simulations of the radio-detection of extensive air showers in GRAND, and to assess the effect of changes to the detector design on simulated observations, while meeting the requirements for simulation production, signal processing, data storage, and data analysis.  To achieve this efficiently, {\sc GRANDlib} interfaces several existing external packages, as described below.

{\sc GRANDlib} uses descriptions of the antenna response and the radio-frequency (RF) chain---the series of electronic components that process and digitize the signal received by an antenna while minimizing background noise---that are specific to GRAND DUs, but has built-in flexibility to accommodate other, similar experiments.

Figure~\ref{fig:grandlib_overview} sketches the GRAND detection principle and the stages of it that {\sc GRANDlib} simulates.  We summarize them below and expand on them later.
\begin{description}
 \item
  [Coordinate systems] The {\tt coordinates} module (Section~\ref{sec:coordinate}) defines terrestrial coordinate systems in which to express antenna positions, the geomagnetic field, and radio signals from air showers, and transformations between them. 
 \item
  [Topography] The {\tt topography} module (Section~\ref{sec:topography}) computes the ground elevation at specified locations on the surface of the Earth. Additionally, via {\sc TURTLE}~\cite{Niess:2019hdn}, it computes the distance traveled by a particle from a specified location along its trajectory before impacting the surface.
 \item
  [Geomagnetic field] The {\tt geomagnet} module (Section~\ref{sec:geomagnet}) computes the strength of the geomagnetic field and its direction at a specified location and time, quantities that are required to model the development of an extensive air shower and its radio emission.
 \item
  [Shower generation] Shower generation (Section~\ref{sec:shower}) is outsourced to simulation tools external to {\sc GRANDlib}. {\sc DANTON}~\cite{Niess:2018opy} handles the simulation of taus made in $\nu_\tau$ interactions underground, and {\sc ZHAireS}~\cite{Alvarez-Muniz:2010hbb} or {\sc CoREAS}~\cite{Huege:2013vt} handles the simulation of extensive air showers and the radio signals from them.
 \item
  [Radio antenna response] The {\tt antenna\_model} and {\tt process\_ant} modules (Section~\ref{sec:ant_response}) implement the simulated response of the GRAND {\sc HorizonAntenna}~\cite{GRAND:2018iaj} for radio waves with frequencies of 30--250~MHZ coming from the upper half-space of the celestial sphere, as expected from horizontal extensive air showers.
 \item
  [Signal processing] The {\tt efield2voltage} module (Section~\ref{sec:signal_processing}) processes the radio signals detected by an antenna and computes the final output voltage.  It uses a model of the antenna response and of the chain of signal-processing electronic components known as the RF chain. The latter is simulated in the {\tt rf\_chain} module.
 \item
  [Data management] The {\tt root\_trees} module (Section~\ref{sec:datamanagement}) defines the data structure used by GRAND simulated and real data and manages their input and output. The {\tt granddb} library facilitates data storage and retrieval.
 \item
  [Data visualization] The {\tt traces\_event} and {\tt du\_network} modules (Section~\ref{sec:visualization}) contains data visualization tools for interactively plotting voltage and electric field traces of an event.  
\end{description}

{\sc GRANDlib} is built on the basis of software developed for TREND~\cite{Ardouin:2010gz}.
It shares some common functionality with simulation software custom-designed for other experiments that detect UHE particles via their radio emission in air~\cite{PierreAuger:2011btp}---like the Pierre Auger Observatory~\cite{PierreAuger:2018pmw}---or in ice \cite{Glaser:2019rxw, Glaser:2019cws}---like RNO-G~\cite{RNO-G:2020rmc} and the planned IceCube-Gen2~\cite{IceCube-Gen2:2020qha}.  
It benefits from experience gained with well-established software packages developed by other experiments, like the IceCube Neutrino Observatory~\cite{DeYoung:2005zz} and the Pierre Auger Observatory~\cite{Argiro:2007qg}. Beyond its primary applications, {\sc GRANDlib} can serve other, similar experiments as a user-friendly tool to convert between shower and array coordinate systems, perform topography-related computations, and query the value of the geomagnetic field at different locations; in addition, it offers a data format optimized for efficient storage and processing.


\section{Coordinate systems}
\label{sec:coordinate}

The {\sc GRANDlib} \texttt{coordinates} module handles the different coordinate systems used by GRAND: Earth-centered Earth-fixed (ECEF), geodetic, Local Tangential Plane (LTP), and GRANDCS, the primary coordinate system used in GRAND.  \ref{appnd1} contains a detailed description of the systems and their transformation from one into another.  All coordinate systems account for the curvature of the Earth.

\begin{lstlisting}[language=Python, caption={\textbf{{\sc GRANDlib} code snippet to convert GRANDCS to different coordinate systems.} GRANDCS is the primary coordinate system used by GRAND.  See Section~\ref{sec:coordinate} for details. More examples are available in the {\sc GRANDlib} repository~\cite{grandlib_repository}, under \texttt{examples/geo/coordinates.ipynb}.}, float=t!, label={lst:coord_example}]
from grand import ECEF, Geodetic, LTP, GRANDCS

# At the GP300 site
geod = Geodetic(latitude=40.98, longitude=93.95, # deg
                height=1267)                     # m
gdcs = GRANDCS(x=10, y=10, z=10, location=geod)  # m

# Convert GRANDCS to ECEF, Geodetic, and LTP
gd2ecef, gd2geod = ECEF(gdcs),Geodetic(gdcs)
gd2ltp = LTP(gdcs, location=gdcs, orientation='NWU')
print('gd2ecef', gd2ecef, 'm')
print('gd2geod', gd2geod) # in deg,deg,m
print('gd2ltp', gd2ltp, 'm')

# This script returns:
# gd2ecef [[-332226.78], [4811543.36], [4161586.64]] m
# gd2geod [[40.98009048], [93.9498818], [1276.9996]]
# gd2ltp  [[0], [0], [0]] m

\end{lstlisting}

GRANDCS is a subclass of LTP where the orientation of the reference frame is pre-defined to be North-West-Up (NWU). The x-axis points towards the local geomagnetic North, the y-axis points 90$^\circ$ west of the x-axis, and the z-axis points upward, perpendicularly to both the x-axis and y-axis. Users define the location of the GRANDCS origin in one of the {\sc GRANDlib} coordinate systems and the time of observation, \texttt{obstime}.
To identify the local magnetic North, the \texttt{geomagnet} module (Section~\ref{sec:geomagnet}) returns the local magnetic field intensity and declination angle at the location of the GRANDCS origin and at the requested \texttt{obstime}.
The detector layout is specified using GRANDCS.  To generate the results in this paper, the GRANDCS origin is placed at the GRANDProto300 site in Dunhuang, China, at latitude 40.95$^\circ$, longitude 93.95$^\circ$, and elevation of \unit[1267]{m} (Fig.~\ref{fig:gp13_site}), at \texttt{obstime} of August 28, 2022.

The geodetic and ECEF coordinate systems are ancillary, mainly used for coordinate transformation and to describe the topography and geomagnetic field.  Coordinates may be expressed in Cartesian and spherical representation, via the classes \texttt{CartesianRepresentation} and \texttt{SphericalRepresentation}.  The ECEF, LTP, and GRANDCS coordinate systems are based on \texttt{CartesianRepresentation}. The geodetic coordinate system is handled by the \texttt{GeodeticRepresentation} class.  The \texttt{SphericalRepresentation} class is used to evaluate the antenna response (Section~\ref{sec:ant_response}) in the direction of shower maximum.  To transform between coordinates systems, instances of one class can be passed as arguments to a different class.

Listing~\ref{lst:coord_example} shows an example of the use of the \texttt{coordinates} module to instantiate the above pre-defined {\sc GRANDlib} coordinate systems and transform between them.


\section{Topography}
\label{sec:topography}

The \texttt{topography} module handles the topography of the Earth, which determines, for instance, where an UHE $\nu_\tau$ could interact underground to make a tau that initiates a shower.  It is based on {\sc TURTLE} (Topographic Utilities for tRansporting parTicules over Long rangEs)~\cite{Niess:2019hdn, turtle_repository}, which provides utilities for simulating the long-range transport of particles through a pre-defined topography. {\sc TURTLE} uses ray tracing to follow particle trajectories; the calculation is sped up by using only the topography neighboring the particle along its trajectory.

The \texttt{topography} module computes the ground elevation at locations on the surface of the Earth from 56$^\circ$S to 60$^\circ$N latitude. It uses NASA Shuttle Radar Topography Mission (SRTM)~\cite{2000EOSTr..81..583F, 2007RvGeo..45.2004F} data with a resolution of 1 arcsecond; along the equator, this corresponds to a resolution of about \unit[30]{m}. Elevation can be computed in reference to the sea level, the ellipsoidal Earth model, or a user-defined LTP frame (Section~\ref{sec:coordinate}).  Limited topographical data is included in the {\sc GRANDlib} installation; users can download more from NASA Earthdata~\cite{nasa_srtm}.

\begin{figure}[t!]
 \centering
 \includegraphics[width=\columnwidth]{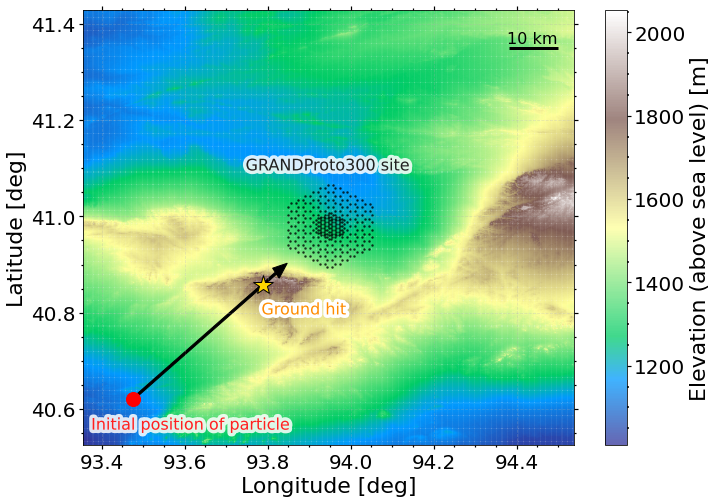}
 \caption{\textbf{Topography of the area surrounding the  GRANDProto300 prototype array in Dunhuang, China.}  The array is centered at 40.98$^\circ$ latitude and 93.95$^\circ$ longitude.
 Overlaid is one of the proposed layouts of GRANDProto300; see Fig.~\ref{fig:data_viz}.
 Locations with surrounding mountains present an ideal topography for GRAND sites. The arrow illustrates the trajectory of a particle starting in air and hitting a nearby mountain; see Fig.~\ref{fig:topo_distance_hit}.
 The topography data used here is from the NASA SRTM mission.
 See Section~\ref{sec:topography} for details. (For interpretation of the colors in the figure(s), the reader is referred to the web version of this article.)}
 \label{fig:gp13_site}
\end{figure}

Figure~\ref{fig:gp13_site} shows, as an example, the topographic map of the area around the GRANDProto300 site. Mountainous regions like in Fig.~\ref{fig:gp13_site} are ideal locations for GRAND, since the mountains act as possible targets for neutrino interactions~\cite{Fargion:1999se, GRAND:2018iaj}.

\begin{figure}[b!]
 \centering
 \includegraphics[width=\columnwidth]{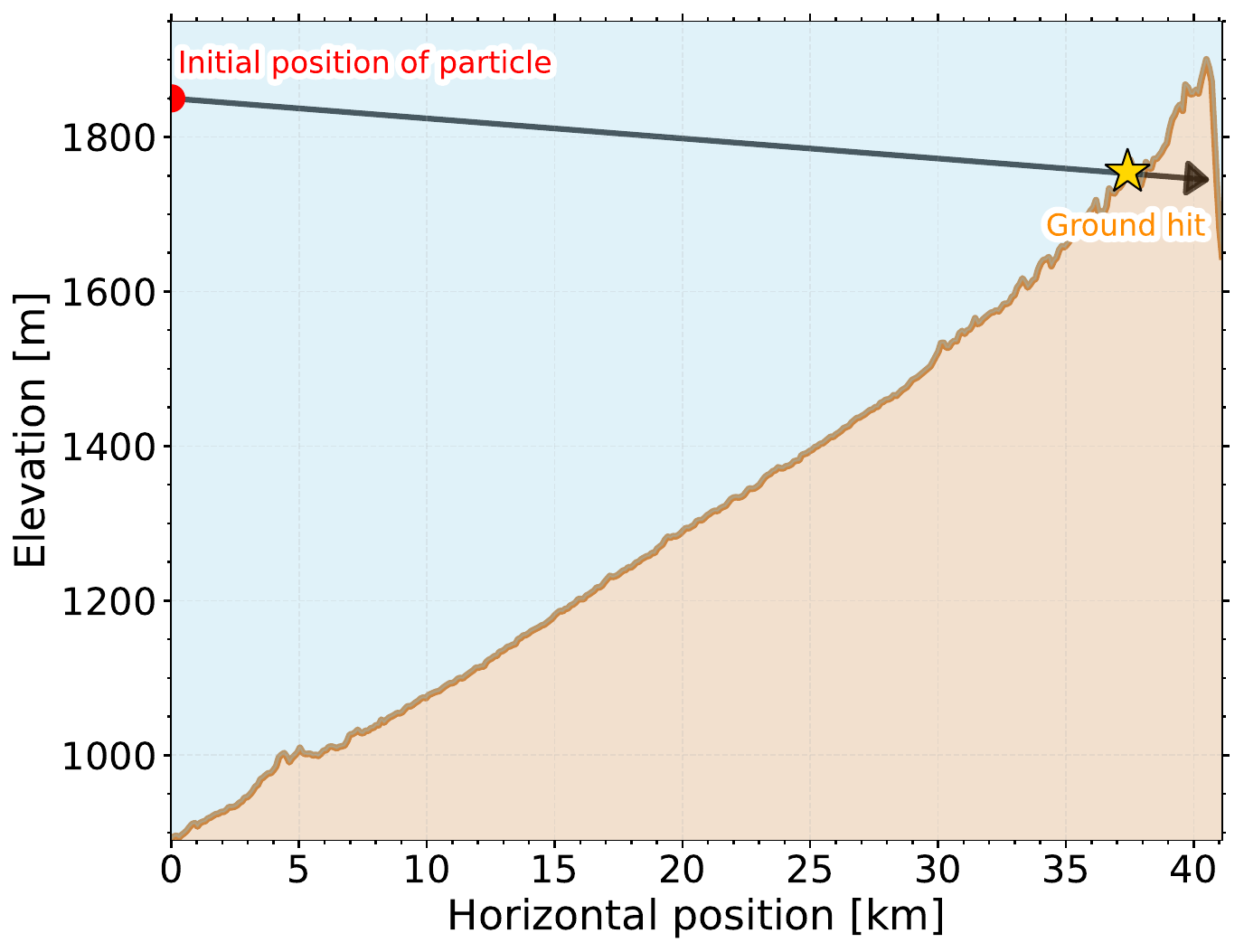}
 \caption{\textbf{Ground elevation tracked along the trajectory of a propagating particle near the GRANDProto300 site.} The trajectory is the same as in Fig.~\ref{fig:gp13_site}. {\sc GRANDlib} computes the distance a particle travels in a specified direction before hitting a nearby mountain. The ground elevation along the particle trajectory is given in GRANDCS coordinates (Section~\ref{sec:coordinate}). Following the trajectory allows {\sc GRANDlib} to determine where Earth-skimming ultra-high-energy $\nu_\tau$ could interact underground. Listing~\ref{lst:topography} shows the code used to calculate the trajectory.  See Section~\ref{sec:topography} for details.}
 \label{fig:topo_distance_hit}
\end{figure}

Figure~\ref{fig:topo_distance_hit} shows the trajectory, calculated with {\sc GRANDlib}, traveled by a particle moving along a near-horizontal direction through the atmosphere before hitting a mountain, like an Earth-skimming UHE $\nu_\tau$ would.  In this case, after hitting the ground, {\sc DANTON}~\cite{Niess:2018opy} computes the neutrino-nucleon deep inelastic scattering, the production of a final-state high-energy tau, its propagation underground and its decay, which initiates an extensive air shower. 

\begin{lstlisting}[language=Python, caption={\textbf{{\sc GRANDlib} code snippet to compute the elevation of a particle traveling in a given topography.}  The particle trajectory is shown in Figs.~\ref{fig:gp13_site} and \ref{fig:topo_distance_hit}.  See Section~\ref{sec:topography} for details. More examples are available in the {\sc GRANDlib} repository~\cite{grandlib_repository}, under \texttt{examples/geo/topography$\_$tutorial.ipynb}.}, float=t!, label={lst:topography}]
from grand import topography
import numpy as np

# GRANDCS at the GP300 site defined asl
geod_asl = Geodetic(latitude=40.98, longitude=93.95,#deg
                height=0)                           # m

# Build a meshgrid for ground topography shown in Fig.2.
bound = np.linspace(-5e4, +5e4, 401)
X, Y  = np.meshgrid(bound, bound)
grnd_grid = GRANDCS(x=X.flatten(), y=Y.flatten(), 
              z=np.zeros(X.size), location=geod_asl)

# Calculate elevation wrt geoid. This is shown in Fig.2.
zg = topography.elevation(grnd_grid, "GEOID")

# Initial point along traj (in GRANDCS ref)
x0 = GRANDCS(x=-4e4, y=4e4, z=1850, location=geod_asl)
# direction vector of a moving particle in GRANDCS ref
direction = GRANDCS(x=10, y=-10, z=-0.036, 
                    location=geod_asl)
# Direction vector must be in ECEF.
dirECEF = np.matmul(x0.basis.T, direction)
dist    = topography.distance(x0, dirECEF)
print("Distance to ground:", dist, "m")

# Build trajectory of a moving particle.
u = np.linspace(0, 1.1*dist, 401) # distance from x0 (m)
# normalize direction vector
dirn = direction / np.linalg.norm(direction)  
traj = dirn * u + x0 
# Get elevation below traj. This is shown in Fig.3.
traj_grand = GRANDCS(x=traj[0], y=traj[1], z=traj[2], location=geod_asl)  # Compute traj coordinates
# z-coordinate of ground in GRANDCS ref
ztG = topography.elevation(traj_grand,reference="LOCAL")  

# This scripts gives:
#   "zg" is the elevation shown in Fig.2.
#   "ztG" is the elevation shown in Fig.3.
#   Distance to ground: 37374.39 m
\end{lstlisting}

Listing~\ref{lst:topography} shows an example of the use of the \texttt{topography} module to generate the particle trajectory in Figs.~\ref{fig:gp13_site} and \ref{fig:topo_distance_hit}.


\section{Geomagnetic field}
\label{sec:geomagnet}

The definition of the GRANDCS system requires as input the declination angle of the local geomagnetic field (Section~\ref{sec:coordinate}).  Further, the intensity of the geomagnetic field, its inclination, and declination angles at a specified location and time are essential inputs to run air-shower simulations (Section~\ref{sec:shower}).  The \texttt{geomagnet} module returns this information via the package {\sc GULL} (Geomagnetic UtiLities Library)~\cite{gull_repository}, which relies on the International Geomagnetic Reference Field (IGRF-13)~\cite{igrf13} and the World Magnetic Model 2020 (WMM2020) models.

In the \texttt{geomagnet} module, the location where the magnetic field should be computed can be provided in two ways: either  specified in one of the coordinate systems defined in Section~\ref{sec:coordinate}, or as \texttt{latitude}, \texttt{longitude}, and \texttt{height}.  The observation time is specified in \texttt{obstime}.  If the geomagnetic field model and \texttt{obstime} are not specified, the default geomagnetic model (`IGRF13') and default \texttt{obstime} (`2020-01-01') are used.

Listing~\ref{lst:geomagnet} shows an example of the use of the \texttt{geomagnet} module to compute the geomagnetic field at the GRANDProto300 site.

\begin{lstlisting}[language=Python, caption={\textbf{{\sc GRANDlib} code snippet to compute the geomagnetic field.}  The field value is returned at the same location and observation time---the GRANDProto300 site (Fig.~\ref{fig:gp13_site}) on August 28, 2022---used to generate the illustrative air-shower simulations and signal processing in Fig.~\ref{fig:test_voltage}. See Section~\ref{sec:geomagnet} for details.  More examples are available in the {\sc GRANDlib} repository~
\cite{grandlib_repository}, under \texttt{examples/geo/geomagnet$\_$tutorial.ipynb}.}, float=t!, label={lst:geomagnet}]
# Geomagnetic field information at the GP300 site used to generate the simulations for this paper.
from grand import Geomagnet

# Accepted formats for `obstime' are `2020-01-01' and datetime.date(2020, 1, 1).
geoB = Geomagnet(location=geod, obstime="2022-08-28")

print("Mag. Field (B):", geoB.field, "Tesla")
print("Observation time:", geoB.obstime)
print("Location:", geoB.location)
print("Declination [deg]:", round(geoB.declination, 3))
print("Inclination [deg]:", round(geoB.inclination, 3))

# This script returns:
#   Mag. Field (B):
#   [[4.4571e-08], [2.6575e-05], [-4.9974e-05]] Tesla
#   Observation time: 2022-08-28
#   Location: [[40.98], [93.95], [1267.0]]
#   Declination [deg]: 0.096
#   Inclination [deg]: 61.997
\end{lstlisting}


\section{Shower generation}
\label{sec:shower}

To simulate extensive air showers triggered by UHE particles and the radio emission from them, {\sc GRANDlib} relies on either {\sc ZHAireS}~\cite{Alvarez-Muniz:2010hbb} or {\sc CoREAS}~\cite{Huege:2013vt}, Monte-Carlo shower simulators that are run separately to {\sc GRANDlib}, and whose output is fed to it.  For details, we defer to Refs.~\cite{Alvarez-Muniz:2010hbb, Huege:2013vt}; below, we merely sketch the process of generating shower simulations.  {\sc ZHAireS} and {\sc CoREAS}  yield similar results~\cite{Huege:2013vu, Gottowik:2017wio, Sanchez:2023vto} and have been shown to match radio measurements~\cite{Nelles:2014xaa, Nelles:2014dja, PierreAuger:2016vya}. 

To generate a shower simulation, the user specifies a few input parameters that define the shower, i.e., the primary particle type, its initial energy, arrival direction---defined by the zenith angle, $\theta$, and the azimuth angle, $\phi$---and a hadronic interaction model with which the particles will interact in the atmosphere (QGSJET-II-04~\cite{Ostapchenko:2010vb} for the results in this paper). The environment in which the shower propagates is characterized by specifying the geomagnetic field amplitude, $B$, its inclination, $i = \arctan{(B_z/B_x)}$, at the shower location, where $B_x$ and $B_z$ are the field components along the x- and z-axes, and a model of the matter density profile of the atmosphere, e.g., the US atmospheric model~\cite{US_atmosphere} or a model generated using the Global Data Assimilation System (GDAS). The simulations used in this paper were generated using an atmospheric model based on GDAS data for a location near the GRANDProto300 site.  The user also provides a list of the antenna positions on the ground at which the electric field of the radio emission from the shower should be computed. 

The main outputs of the shower simulations are the electric field time-traces, $E_x(t)$, $E_y(t)$, and $E_z(t)$, at each antenna position, the longitudinal particle distributions, and the lateral particle distribution at ground level, or at specified altitudes.  The electric fields are stored in the format discussed in Ref.~\cite{Piotrowski:2023xbc} and \ref{appnd:fileformat}.  The resolution of the simulation depends on the value of the thinning parameter, which determines how many secondary particles are tracked during shower development~\cite{Alvarez-Muniz:2010hbb, Huege:2013vt}.  For the simulations in this paper, we used a value of $10^{-5}$, slightly larger than the typical value of $10^{-6}$, which speeds up the simulations.
Figure~\ref{fig:test_voltage} (top panel) shows the resulting electric field from an illustrative simulated shower.

To generate the main results in this paper, shown in Fig.~\ref{fig:test_voltage}, we have used a shower initiated by a proton with an energy of \unit[3.98]{EeV} and zenith angle of 85$^\circ$ simulated by {\sc ZHAireS} at the location of the GRANDProto300 site (Fig.~\ref{fig:gp13_site}) on August 28, 2022.  (Later, in Section~\ref{sec:signal_processing-runtime_benchmark}, we benchmark the run times of {\sc GRANDlib} using a range of proton energies and zenith angles.)


\section{Radio antenna response}
\label{sec:ant_response}

\begin{figure}[t!]
 \centering
 \includegraphics[trim={0 0.5cm 0 0.85cm}, clip, width=0.75\columnwidth]{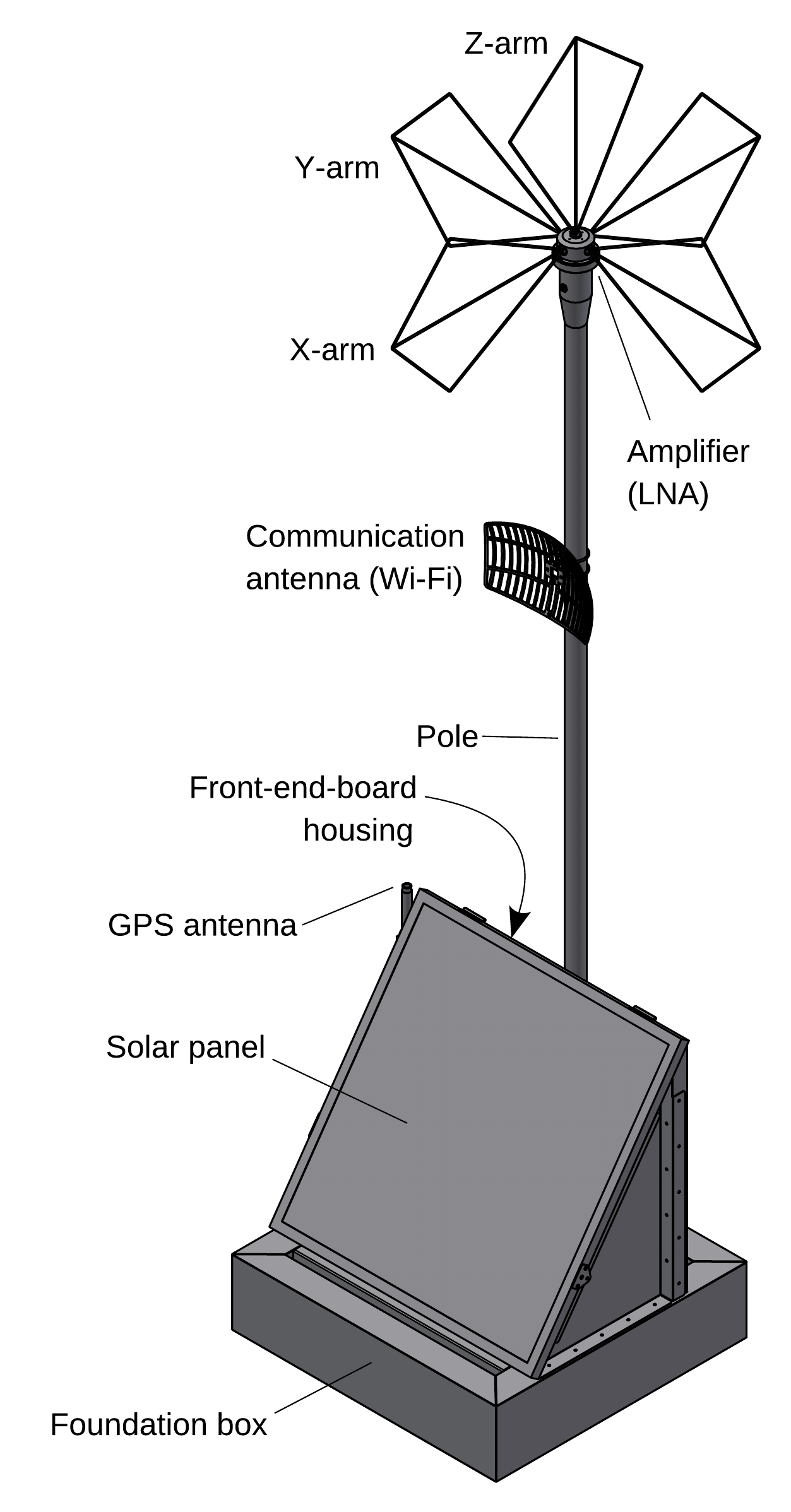}
 \caption{\textbf{Schematic diagram of a GRANDProto300 detection unit.} {\sc GRANDlib} models the response of the X-, Y-, and Z-arms of the GRAND {\sc HorizonAntenna} to 
 the radio signal from an extensive air shower (Section~\ref{sec:ant_response}) and the subsequent signal processing of the recorded voltage (Section~\ref{sec:signal_processing}), which takes place in the LNA and the front-end board.  The foundation box is filled with sand to weigh down the structure.}
 \label{fig:antenna}
\end{figure}

\begin{figure*}[t!]
 \centering
 \includegraphics[width=\columnwidth, trim={0.1cm 0.1cm 0 1.4cm}, clip=False]{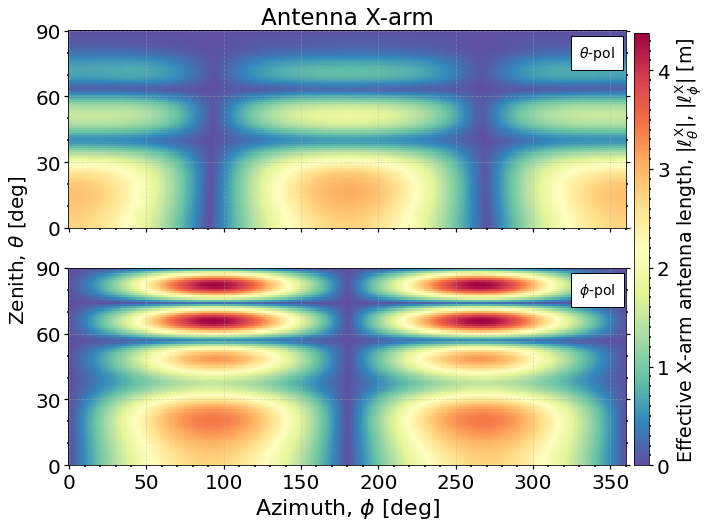}
 \includegraphics[width=\columnwidth, trim={0.1cm 0.1cm 0 1.4cm}, clip=False]{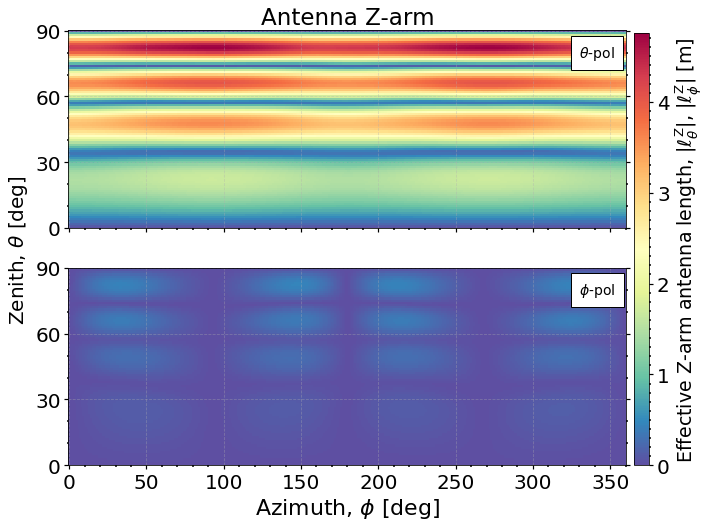}
 \caption{\textbf{Simulated magnitude of the effective length of GRAND {\sc HorizonAntenna}.}  The effective antenna length represents the response of the antenna to an incoming radio signal. In this plot, the response is shown at \unit[150]{MHz}, in the local spherical coordinate system of an antenna, in the $\hat{\boldsymbol{\theta}}$ (\textit{top}), and $\hat{\boldsymbol{\phi}}$ (\textit{bottom}) polarizations. The component in the $\hat{\boldsymbol{r}}$-polarization (not shown) is negligible because of the null strength of the electric field along its direction of propagation.  \textit{Left:} Effective length of the X-arm of the antenna.  The effective length of the Y-arm (not shown) is similar, but shifted $90^\circ$ in azimuth relative to the X-arm, since the X- and Y-arms are perpendicular to each other (Fig.~\ref{fig:antenna}).  \textit{Right:} Effective length of the Z-arm.   See Section~\ref{sec:ant_response} for details.}
 \label{fig:leff}
\end{figure*}

Figure~\ref{fig:antenna} shows one of the GRANDProto300 DUs, representative of the design of GRAND DUs.  In what follows, we focus our discussion on this DU design, but the GRAND@Auger and GRAND@Nan\c cay DUs are similar~\cite{deMelloNeto:2023zvk} and can also be simulated by {\sc GRANDlib}.  Each DU has three mutually perpendicular antenna arms arranged in South-North (SN), East-West (EW), and vertical (Z) directions, where the local geomagnetic north is used as the North direction~\cite{ARENAKotera24}. The antenna arms are referred to as the X-, Y-, and Z-arm, respectively. Each arm is composed of a symmetric butterfly-shaped steel radiator. The X- and Y-arms are dipoles, whereas the Z-arm direction is a monopole, with the ground and the supporting pole acting as its second arm; below, we show that this results in marked differences in the response of the X- and Y-arms \textit{vs.}~the Z-arm.

The electric field incident from the upper half-space, propagating along zenith angle $\theta$ and azimuth angle $\phi$, is $\boldsymbol{E}(\theta, \phi, f) = E_x \hat{\boldsymbol{x}} + E_y \hat{\boldsymbol{y}} + E_z \hat{\boldsymbol{z}}$, where $f$ is the frequency of the antenna response, $\hat{\boldsymbol{x}}$, $\hat{\boldsymbol{y}}$, and $\hat{\boldsymbol{z}}$ are unit basis vectors in a given coordinate system, and $E_x$, $E_y$, and $E_z$ are the components, or polarizations, of the field.  [The direction of the incoming electric field varies for different DUs depending on their location. To calculate the direction at each DU, we define a LTP frame (Section~\ref{sec:coordinate}) for each of them, and transform the position of the shower maximum---from where we assume the electric field is emitted---to the LTP frame of each DU.]

The response of an antenna to the incoming electric field from an air shower is contained in the antenna effective length~\cite{balanis}, a vector quantity that characterizes the ability of the antenna to detect an incident electric field and convert it into a voltage, which is afterward processed (Section~\ref{sec:signal_processing}).  It is represented as $\boldell^p
(\theta, \phi, f) = \ell_x^p \hat{\boldsymbol{x}} + \ell_y^p \hat{\boldsymbol{y}} + \ell_z^p \hat{\boldsymbol{z}}$ in Cartesian coordinates and $\boldell^p(\theta, \phi, f) = l_\theta^p \hat{\boldsymbol{\theta}} + l_\phi^p \hat{\boldsymbol{\phi}}$ in spherical coordinates, where $p 
= X, Y, Z$ denotes each antenna arm, whose voltage is read in the corresponding X-, Y-, or Z-port.  (The $r$-polarization of the effective length is negligible, since the electric field is small in the direction parallel to its propagation.  In {\sc GRANDlib}, it is set to zero.)
The resulting \textit{open-circuit voltage} (Section~\ref{sec:signal_processing-open_circuit_voltage}) is a complex quantity with an amplitude and phase that quantify the ability of an antenna to respond to the three polarizations of $\boldsymbol{E}$.

In {\sc GRANDlib}, the effective length of each of the antenna arms of the GRAND {\sc HorizonAntenna}~\cite{GRAND:2018iaj}---designed to be sensitive to very inclined air showers---is stored as a pre-computed table for the X-, Y-, and Z-arms, for $f \in [30, 250]$~MHz, $\theta \in [0^\circ, 90^\circ]$, and $\phi \in [0^\circ, 360^\circ]$, and is interpolated for values that are not pre-computed, its real and imaginary parts treated independently.  The effective length accounts for reflections of the radio signals off of the ground, and for the height of the antenna pole, of 3.2~m in the {\sc HorizonAntenna}.  The effective length used in {\sc GRANDlib} holds regardless of whether the radio signal is short and broadband (i.e., wide in its frequency range), or long and narrowband.

Figure~\ref{fig:leff} shows the magnitude of the effective length for the X- and Z-arms of a {\sc HorizonAntenna} at a frequency of \unit[150]{MHz}, for an incoming electric field along different directions.  This frequency is representative of the shape of the structures in the effective length, but the structures change in size and number for other frequencies; this is accounted for in {\sc GRANDlib}.
The effective length is maximum when a component of the electric field oscillates parallel to the antenna arm. 

In Fig.~\ref{fig:leff}, for the X-arm (and also for the Y-arm, not shown), the $\phi$-polarization of the electric field oscillates parallel to the X-arm of the antenna when the shower direction is $\phi = 90^\circ$ or $\phi = 270^\circ$. This results in the effective length being maximum along the $\phi$-polarization where it is minimum for the $\theta$-polarization.  For the Z-arm, because it is vertical (Fig.~\ref{fig:antenna}), its effective length is almost exclusively in the $\theta$-polarization.  The striped pattern in the $\theta$-polarization is due to the destructive interference of the radio signal with itself after reflecting off the ground. The reflection lengthens the path of the signal, delaying its arrival to the antenna by up to tens of ns, with the exact delay being different for different frequencies (and for different incidence angles of the signal), and the net delay being due to the interference of the different frequencies.  A detailed study of the antenna effective length will be presented elsewhere.


\section{Signal processing}
\label{sec:signal_processing}

One of the main goals of {\sc GRANDlib} is the mass production of end-to-end simulations encompassing shower generation, radio-signal detection, and signal processing (Fig.~\ref{fig:grandlib_overview}).

First, we use {\sc ZHAireS} or {\sc CoREAS} to simulate the electric fields of radio signals generated in extensive air showers.  Second, at each DU, the electric field is detected by a passive antenna and converted into an open-circuit voltage signal using the antenna effective length (Section~\ref{sec:ant_response}).  We then add noise---of Galactic and anthropogenic origin---to the open-circuit voltage. Finally, the open-circuit voltage is transmitted through various components of the RF chain (Section~\ref{sec:signal_processing-rf_chain}) of the DU to generate a final output voltage, which is then digitized by an analog-to-digital converter (ADC). By default, digitization is carried out with 14-bit precision at a sampling rate of 500 million samples per second and the conversion factor is 1~ADC equivalent to 109.86~$\mu$V~\cite{Ma:2023siw}; these values can be adjusted in the code. The digitization stores traces from four channels, each with a trace length of 4096 bins of \unit[2]{ns}, with each bin requiring 2 bytes of storage.  Three channels store data from the X-, Y-, and Z-arms of the antenna; the fourth channel typically collects noise, and is used for testing and as backup.

Below, we outline how {\sc GRANDlib} simulates the conversion of an input electric field into an output voltage.  


\subsection{Open-circuit voltage}
\label{sec:signal_processing-open_circuit_voltage}

We express the incoming electric field, $\boldsymbol{E}$, in the frequency domain via a fast Fourier transform, within the 30--250~MHz range.  The response of the antenna to the electric field generates an open-circuit voltage,
\begin{equation}
 V_{\rm oc}^p
 =
 \boldell^p \cdot \boldsymbol{E}
 =
 \ell_x^p E_x + \ell_y^p E_y + \ell_z^p E_z \;,
 \label{eqn:voc}
\end{equation}
where $\boldell^p$ is the antenna effective length  (Section~\ref{sec:ant_response}), and, like before, $p$ denotes each of the three antenna arms.

\begin{figure}[t!]
 \centering
 \includegraphics[width=\columnwidth]{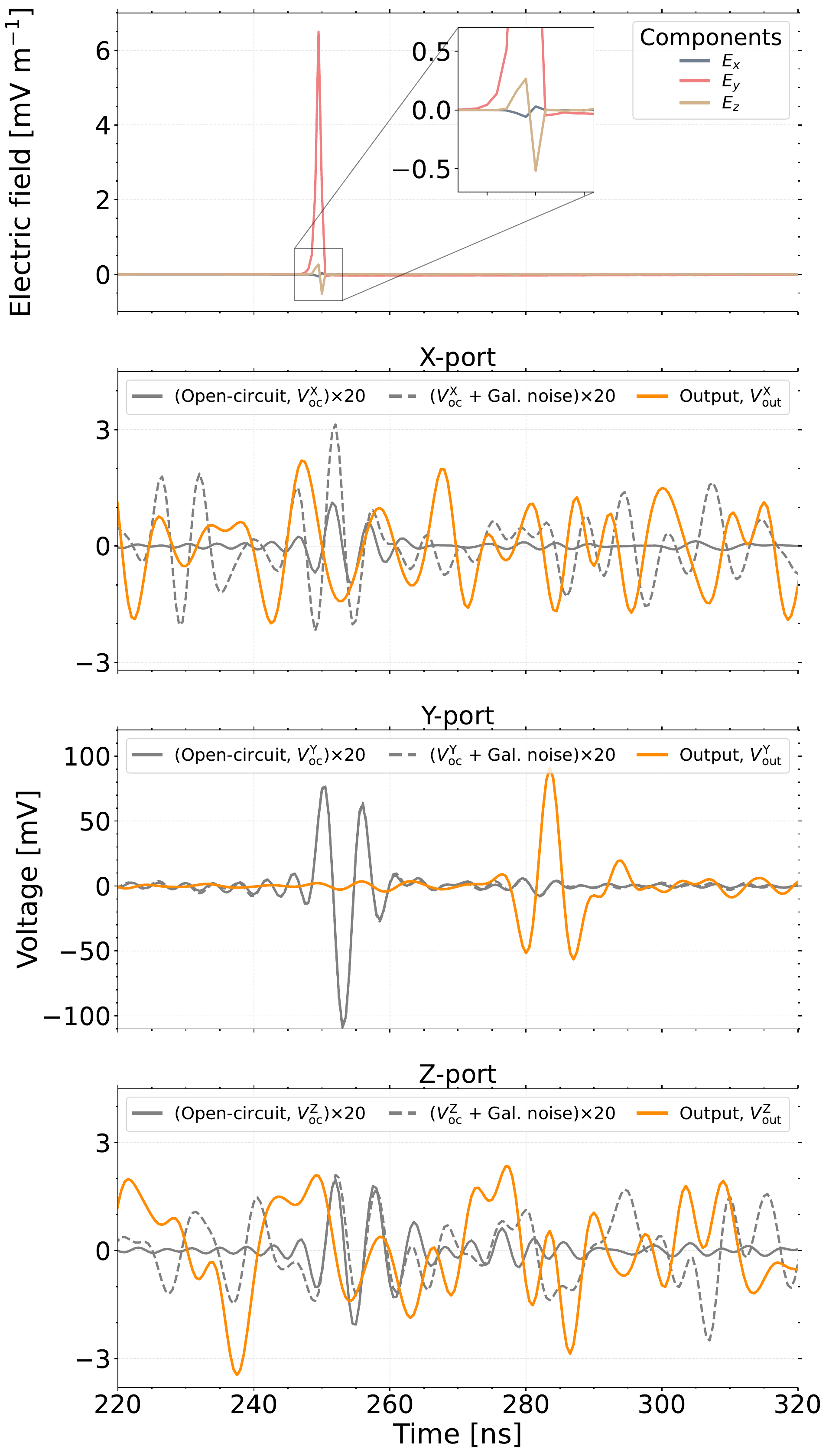}
 \caption{\textbf{Simulated end-to-end processing of the radio signal of an extensive air shower by a GRAND detection unit, via {\sc GRANDlib}.}
 \textit{Top:} Components of the electric field generated in a simulated extensive air shower initiated by a cosmic-ray proton of energy \unit[3.98]{EeV} with a zenith angle of 85$^\circ$ and azimuth angle of 0$^\circ$, occurring at LST = 18 hours. The electric field is computed at ground level, at the location of an illustrative GRAND DU on the GRANDProto300 site (Fig.~\ref{fig:gp13_site}).  \textit{Bottom three panels:} Simulated open-circuit voltage, Eq.~(\ref{eqn:voc}), computed for this electric field, with and without additional Galactic noise, and output voltage, obtained after processing it through the RF chain of the DU using a 20-dB VGA gain.  The voltage on each antenna arm is measured at the output port of that arm.  See Sections~\ref{sec:signal_processing-open_circuit_voltage} and \ref{sec:signal_processing-output_voltage} for details.}
 \label{fig:test_voltage}
\end{figure}

\begin{figure}[t!]
 \setlength{\lineskip}{12pt}
 \centering
 \includegraphics[width=\columnwidth]{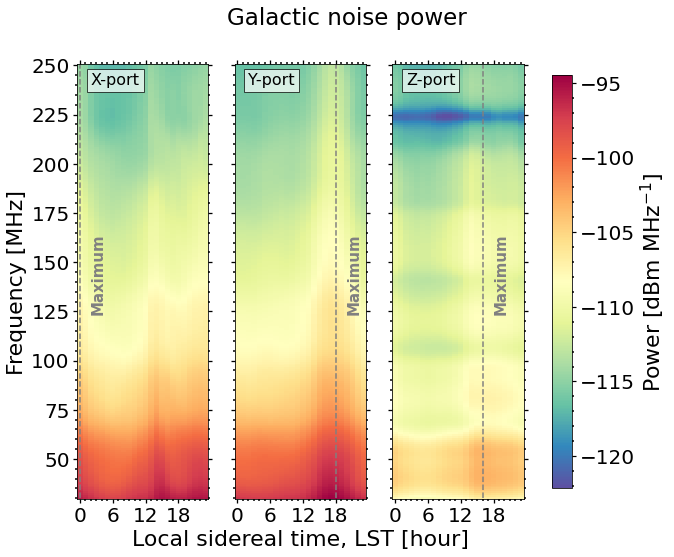}\\*
 \includegraphics[width=\columnwidth]{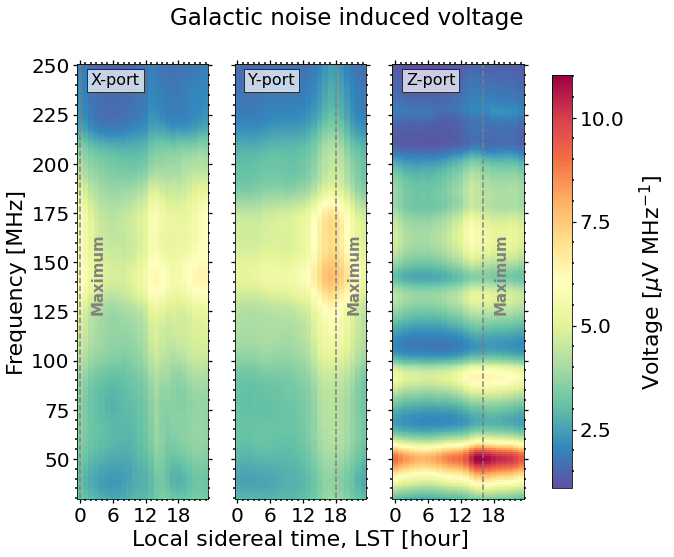}
 \caption{\textbf{Simulated Galactic radio noise induced in a GRAND {\sc HorizonAntenna} in GRANDProto300.}
 The location is in Dunhuang, China (latitude of 40.98$^\circ$, longitude of 93.95$^\circ$); see Fig.~\ref{fig:gp13_site}.  \textit{Top:} Spectral power distribution.  The Galactic noise is computed using \texttt{LFMap}.  The noise is in ports X (\textit{left}), Y (\textit{middle}), and Z (\textit{right}). The maximum Galactic-noise power across all frequencies occurs at LST = 24, 18, and 16 hours, for ports X, Y, and Z, respectively.  \textit{Bottom:} Voltage induced in the antenna arms by the Galactic noise.  The voltage is computed by multiplying the electric field of the Galactic noise (not shown) times the effective antenna length (Fig.~\ref{sec:ant_response}).  See Section~\ref{sec:signal_processing-galactic_noise} for details.}
 \label{fig:gal_noise}
\end{figure}

Figure~\ref{fig:test_voltage} (top panel) shows the components of a sample simulated electric field from a proton-initiated air shower at an illustrative DU. Showers initiated by cosmic rays generate impulsive, nanosecond-length electric fields in radio frequencies~\cite{Jelley1966, Ardouin:2010gz, deVries:2011pa, Schellart:2014oaa, Huege:2016veh}. Due to the refractive index of the atmosphere, most of the signal is concentrated within a cone-shaped geometry~\cite{Schroder:2016hrv, Huege:2017khw}.
Figure~\ref{fig:test_voltage} (bottom three panels) also shows the open-circuit voltage induced in the three antenna arms by the electric field. The formerly sharp electric field signal is smeared and elongated in time due to the antenna response.


\subsection{Galactic noise}
\label{sec:signal_processing-galactic_noise}

To compute the Galactic radio noise, {\sc GRANDlib} uses {\sc LFMap}~\cite{lfmap}, which returns the brightness temperature across the sky, which included the cosmic microwave background, an isotropic component, and an anisotropic component from the Milky Way.  The response of a GRAND {\sc HorizonAntenna} to Galactic noise is pre-computed and tabulated for frequencies of 30--250~MHz and local sidereal times (LSTs) of 0--24~hours.  Using it, the \texttt{galaxy} module returns the voltage induced by the Galactic noise induced at each antenna arm, at requested values of frequency and LST. 

{\sc GRANDlib} does not account for the spatial coherence of the Galactic noise registered by neighboring DUs throughout the array, which is expected to be low due to the sparsity of the array.  Instead, the \texttt{galaxy} module mimics the variability of the Galactic noise by multiplying, for each DU, the sky-averaged value of the noise by a different random value of its phase for each antenna arm and at each frequency.

Figure~\ref{fig:gal_noise} (top panel) shows the power spectral density generated by Galactic noise in a {\sc HorizonAntenna} across various frequencies and times at the GRANDProto300 site (Fig.~\ref{fig:gp13_site}).  
The antenna arms are exposed to the maximum Galactic noise at times when the array site faces the Galactic Center.  Seen from the GRANDProto300 site, the Galactic Center is closer to the horizon than to the zenith; as a result, the X- and Y-arms of the antennas are exposed to more Galactic noise than the Z-arm, the only arm with vertical orientation (Fig.~\ref{fig:antenna}).  A full description of the {\sc HorizonAntenna} response, and of its sensitivity to the Galactic noise, lies beyond the scope of this paper and will be presented elsewhere.

Figure~\ref{fig:gal_noise} (bottom panel) also shows the corresponding voltage induced in the antenna by Galactic noise.  For the X- and Y-arms, the induced voltage is similar because their Galactic power spectral density and effective antenna lengths (Fig.~\ref{fig:leff}) are similar.  For the Z-arm, the induced voltage is markedly different, since the effective antenna length is, too.  The Galactic-noise voltage in the Z-arm exhibits stripes of low and high values at different frequencies, a feature inherited from the stripes of sensitivity that the effective antenna length has at different zenith angles, which are visible in Fig.~\ref{fig:leff} for 150~MHz, but exist also at other frequencies.  Notably, at low frequencies, around 50~MHz, the voltage induced in the Z-arm is appreciably larger than in the X- and Y-arms.

Figure~\ref{fig:test_voltage} shows the effect of adding the Galactic-noise voltage to the open-circuit voltage of a simulated air shower detected by a GRAND DU.  In the X- and Y-arms, the induced Galactic-noise voltages are similar in absolute size---as expected from Fig.~\ref{fig:gal_noise}---but the relative effect of the noise is more prominent in the X-arm, where its magnitude is comparable to that of the open-circuit voltage from the air shower.  In the Z-arm, the magnitude of the induced Galactic-noise voltage is comparable to that of the X- and Y-arms, even though the Galactic-noise voltage is smaller (Fig.\ref{fig:gal_noise}).  Figure~\ref{fig:gal_noise} reveals that this is a consequence of the large boost in Galactic noise in the Z-arm around 50~MHz, which compensates for the smaller Galactic-noise voltage at higher frequencies. 


\subsection{Radio-frequency (RF) chain}
\label{sec:signal_processing-rf_chain}

\begin{figure}[t!]
 \centering
 \includegraphics[width=\columnwidth]{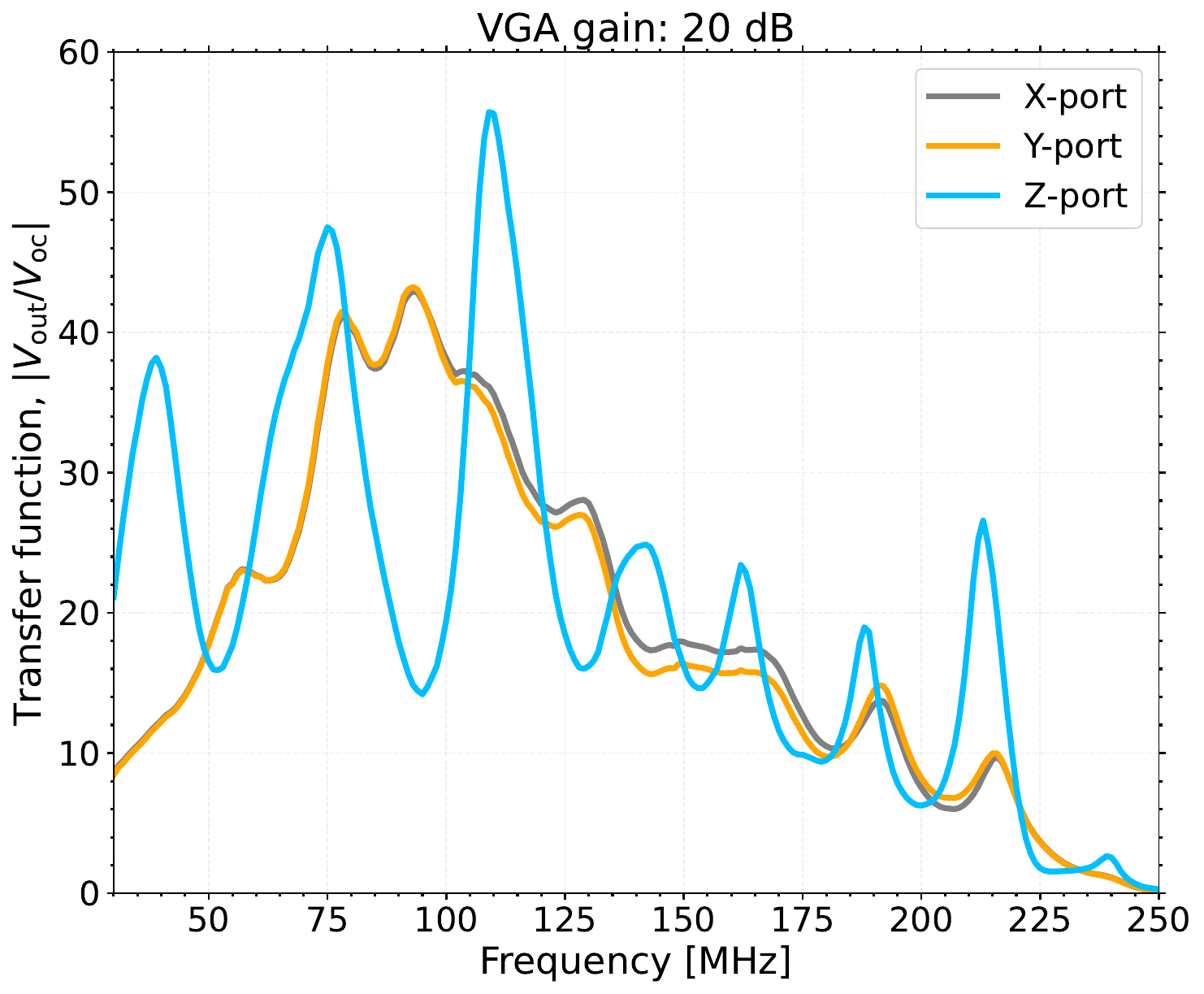}
 \caption{\textbf{Simulated modulus of the total transfer function of the RF chain of a GRAND DU.}  In this plot, the voltage amplification is for an illustrative choice of \unit[20]{dB} VGA gain, computed using {\sc GRANDlib}.  The total transfer function quantifies the net effect of the RF chain on the open-circuit voltage (Section~\ref{sec:signal_processing-open_circuit_voltage}).
 For the Z-port, the antenna pole functions as an antenna arm (Fig.~\ref{fig:antenna}, yielding a transfer function different from that of the X- and Y-ports.  See Section~\ref{sec:signal_processing-rf_chain} for details.}
 \label{fig:total_tf}
\end{figure}

The radio-frequency receiving and processing system of a GRAND DU consists of passive antenna probes, an RF chain, and an ADC system. The passive antenna probe consists of an antenna arm, from among the X-, Y-, and Z-arms. The RF chain consists of a low-noise amplifier (LNA), a cable with a connector, a variable-gain amplifier (VGA), and a filter. The ADC sampling system consists of a balun, an internal 200-$\Omega$ load, and an ADC chip. The \texttt{rf\_chain} module computes the response of each component of the RF chain and the ADC system using their measured scattering parameters (S-parameters). The S-parameters quantify the reflection and transmission of radio signals as they travel through the components of the DU.

The \texttt{rf\_chain} module calculates the total transfer function, $\vert V_{\rm out}^p / V_{\rm oc}^p \vert$, which quantifies how the RF chain modifies the initial voltage signal received by the antennas into the output voltage, $V_{\rm out}^p$ (Section~\ref{sec:signal_processing-output_voltage}). The total transfer function changes with the choice of VGA gain; the output voltage grows with the VGA gain. {\sc GRANDlib} includes S-parameters for VGA gains of \unit[20]{dB}, \unit[5]{dB}, \unit[0]{dB}, and \unit[-5]{dB}. 

Figure~\ref{fig:total_tf} shows the total transfer function of the RF chain for a 20-dB VGA gain---the default value used in GRANDProto300---for the three antenna ports. The total transfer function for the Z-port is different from that of the X- and Y-ports because of different RF chain setups~\cite{grand_prototype_paper}.


\subsection{Output voltage}
\label{sec:signal_processing-output_voltage}

Figure~\ref{fig:test_voltage} shows the final output voltage, $V_{\rm out}^p$, for a GRANDProto300 DU.  It is computed as the product, in the frequency domain, of the open-circuit voltage times the total transfer function shown in Fig.~\ref{fig:total_tf}, after which the final output voltage is transformed back into the time domain via an inverse fast Fourier transform.  The time offset between the output and open-circuit voltages in Fig.~\ref{fig:test_voltage} is due to the delay introduced by processing the signal through the RF chain.


\subsection{Run-time benchmarks}
\label{sec:signal_processing-runtime_benchmark}

Listing~\ref{lst:sig_processing} shows an example of how to compute the output voltage from a given electric field.

To benchmark the run time of {\sc GRANDlib} to do this, we use 300 simulated showers initiated by protons with energies of 0.2--3.98~EeV, zenith angles of 70$^\circ$--85$^\circ$, and azimuth angles of 0$^\circ$, 90$^\circ$, 180$^\circ$, and 270$^\circ$.  Like before, the showers were simulated at the GRANDProto300 site (Fig.~\ref{fig:gp13_site}) on August 28, 2022, using {\sc ZHAireS}.  We convert the electric field of each shower into an output voltage, as described above.

On average, the time required by {\sc GRANDlib} to convert the electric field of a shower into an output voltage in all the DUs that make up GRANDProto300 is about \unit[13]{s}, when running it on a single core of a personal computer with a 1.3-GHz Intel Core i5 processor. This run time is small compared to the time required to simulate an extensive air shower and the radio emission from it, which can take several hours.

\begin{lstlisting}[language=Python, caption={\textbf{{\sc GRANDlib} code snippet to compute the output voltage in a GRAND detection unit.} The incident  electric field is pre-computed from a simulated extensive air shower.  {\sc GRANDlib} adds to it Galactic noise, and processes the result through the RF chain to generate the output voltage.  See Section~\ref{sec:signal_processing} for details.  More details are available in the {\sc GRANDlib} repository~\cite{grandlib_repository}, under \texttt{scripts/convert$\_$efield2voltage.py}.}, float=t!, label={lst:sig_processing}]
# An example script to compute voltage from electric field.
from grand import Efield2Voltage

signal = Efield2Voltage("input_efield.root", "output_voltage.root")
signal.params["add_noise"]    = True
signal.params["add_rf_chain"] = True

#Computed voltage is stored automatically in "output_voltage.root"
signal.compute_voltage()
\end{lstlisting}


\section{Data management}
\label{sec:datamanagement}

GRAND uses a multi-level trigger system.  In the first-level trigger, the local data acquisition (DAQ) system of each DU stores the detected output voltages~\cite{Xu:2023mpw, LeCoz:2023bie}.  The occurrence of a first-level trigger is communicated by the DU to a central DAQ system; data transfer between the DU and the central DAQ system is via a dedicated communications antenna (Fig.~\ref{fig:antenna}).  In the central DAQ system, a second-level trigger assesses whether the data stored locally at the DU should be transmitted to the central DAQ system for permanent storage and eventual analysis; see Refs.~\cite{Correa:2023maq, Mitra:2023pha, Ma:2023siw, Duan:2023rtd} for preliminary work.  

A prominent feature of {\sc GRANDlib} is its ability to handle large volumes of simulated and experimental data. Efficient storage and processing requires a data format that is optimized for input-output speed and compression. The CERN ROOT~\cite{Brun:1997pa} \texttt{TFile} and \texttt{TTree} formats have been selected for this purpose. A comprehensive description of the data format used in {\sc GRANDlib} is provided in Ref.~\cite{Piotrowski:2023xbc} and \ref{appnd:fileformat}. The \texttt{root\_trees} module handles data formatting.

GRAND data are registered in an official database at the CC-IN2P3 computing center in Lyon, France, and are stored in one or more official repositories. Prior to that, they are verified for correctness in structure, data format, naming conventions, etc.

The \texttt{granddb} library provides a mechanism for users to create a local read-write replica of the official GRAND database, and to keep it synchronized with it.
The \texttt{granddb} library also handles data storage and retrieval. Users can search for files by filename or by associated metadata. If the requested files are present in the local database, \texttt{granddb} returns handles to these files. If the files are not found locally, \texttt{granddb} retrieves them from the remote repositories where they are stored, copies them into the local directory, and returns the handles to the local files.


\section{Data visualization}
\label{sec:visualization}

{\sc GRANDlib} has a built-in interactive visualization tool to display the distribution of DU hits from an air shower, and the associated electric fields, voltage traces, and power spectra.
Given an input file, the visualization tool can generate several types of plots.
[Beside plotting, it can also retrieve a list of all DUs and \texttt{TTrees} (\ref{appnd:fileformat}) in the input file.]

\begin{figure}[t!]
 \centering
 \includegraphics[width=\columnwidth]{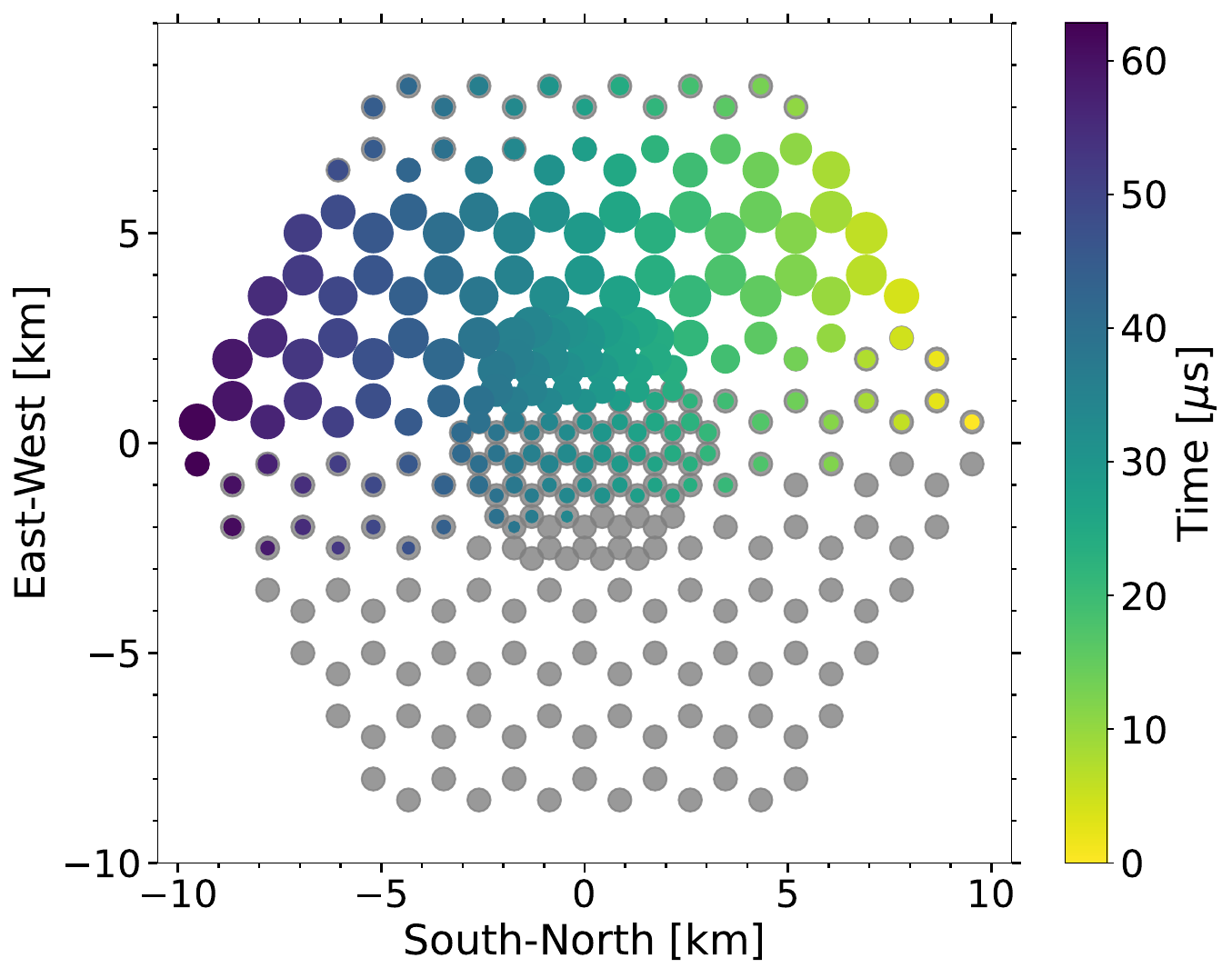}
 \caption{\textbf{Simulated arrival times at GRAND DUs of the radio signal emitted by an extensive air shower.} The shower is initiated by a proton of energy 3.98~EeV and zenith angle of $85^\circ$, at a location near Dunhuang, China. The plot is produced by the {\sc GRANDlib} data visualization tool (using a version presently under development). Each circle represents a DU. Color represents the time at which the DU is hit by the radio signal from the shower; the size of the circle represents roughly the signal strength deposited by the radio signal on that DU.  Gray circles are DUs that are not hit. The layout is one of the proposed alternatives under consideration for GRANDProto300.  See Section~\ref{sec:visualization} for details.}
 \label{fig:data_viz}
\end{figure}

Figure~\ref{fig:data_viz} shows an example plot displaying a proposed GRANDProto300 layout, the times at which the radio signal from a simulated air shower hits each DU, and the scaled peak amplitude of the electric field on each DU accumulated over time. When running this plot interactively in {\sc GRANDlib}, clicking on a DU displays the time traces and power spectra in it.

Additional plotting options include a snapshot of DU hits across the detector layout at user-selected times during the development of the air shower, and time traces and power spectra of any chosen DU.


\section{Summary and outlook}
\label{sec:summary}

{\sc GRANDlib}~\cite{grandlib_repository} is an open-source software tool designed to facilitate the development and data analysis of an experiment---such as GRAND---whose goal is to detect the radio emission from extensive air showers triggered by ultra-high-energy cosmic rays, gamma rays, and neutrinos.

{\sc GRANDlib} serves as the framework for developing simulation and analysis code for GRAND. It comprises tools to manage the end-to-end computation of the detection of radio signals from ultra-high-energy particles, ranging from the physics of particle interactions and showers---including terrestrial coordinate systems, geomagnetism, and topography---to the detection and processing of the radio signals they trigger---including the antenna response, radio-frequency chain, and Galactic and anthropogenic noise.  {\sc GRANDlib} manages the storage and retrieval of experimental and simulated data in remote and local databases, and can generate different types of plots with them.

{\sc GRANDlib} is in active development by the GRAND Collaboration to meet the evolving needs of the experiment.  Planned improvements include a smoother interface with {\sc ZHAireS} and {\sc CoREAS}, more advanced visualization tools, and tools to streamline the processing pipeline and data analysis.  There is also ongoing work in event reconstruction~\cite{Kohler:2024oqm, GRAND:2024tdn}, i.e., inferring properties such as the shower energy and direction, and the identity of the primary particle, from the detected radio signals.

Although {\sc GRANDlib} is tailored to the needs of the GRAND experiment, it can be adopted, in full or partially, by other radio-detection experiments by modifying detector specifications.

As the volume of experimental and simulated data generated by GRAND and other radio experiments grows, {\sc GRANDlib} will provide the backbone needed to analyze them with realistic experimental nuance.


\section{Acknowledgements}
\label{sec:acknowledgements}

The GRAND Collaboration is grateful to Arno Engels, from Radboud University Nijmegen, for providing the diagram of the GRAND {\sc HorizonAntenna} in Fig.~\ref{fig:antenna}, and to Fei Gao, from Xidian University, for their work on antenna design.  The GRAND Collaboration acknowledges the support from the following funding agencies and grants.
\textbf{Brazil}: Conselho Nacional de Desenvolvimento Cienti\'ifico e Tecnol\'ogico (CNPq); Funda\c{c}ão de Amparo \`a Pesquisa do Estado de Rio de Janeiro (FAPERJ); Coordena\c{c}ão Aperfei\c{c}oamento de Pessoal de N\'ivel Superior (CAPES).
\textbf{China}: NAOC, National SKA Program of China (Grant No.~2020SKA0110200); Project for Young Scientists in Basic Research of Chinese Academy of Sciences (No.~YSBR-061); Program for Innovative Talents and Entrepreneurs in Jiangsu, and High-end Foreign Expert Introduction Program in China (No.~G2023061006L); China Scholarship Council (No.~202306010363).
\textbf{Denmark}: Villum Fonden (project no.~29388).
\textbf{France}: ``Emergences'' Programme of Sorbonne Universit\'e; France-China Particle Physics Laboratory; Programme National des Hautes Energies of INSU; for IAP---Agence Nationale de la Recherche (``APACHE'' ANR-16-CE31-0001, ``NUTRIG'' ANR-21-CE31-0025, ANR-23-CPJ1-0103-01), CNRS Programme IEA Argentine (``ASTRONU'', 303475), CNRS Programme Blanc MITI (``GRAND'' 2023.1 268448), CNRS Programme AMORCE (``GRAND'' 258540); Fulbright-France Programme; IAP+LPNHE---Programme National des Hautes Energies of CNRS/INSU with INP and IN2P3, co-funded by CEA and CNES; IAP+LPNHE+KIT---NuTRIG project, Agence Nationale de la Recherche (ANR-21-CE31-0025); IAP+VUB: PHC TOURNESOL programme 48705Z. 
\textbf{Germany}: NuTRIG project, Deutsche Forschungsgemeinschaft (DFG, Projektnummer 490843803); Helmholtz—OCPC Postdoc-Program.
\textbf{Poland}: Polish National Agency for Academic Exchange within Polish Returns Program no.~PPN/PPO/2020/1/00024/U/00001,174; National Science Centre Poland for NCN OPUS grant no.~2022/45/B/ST2/02889.
\textbf{USA}: U.S. National Science Foundation under Grant No.~2418730.
Computer simulations were performed using computing resources at the CCIN2P3 Computing Centre (Lyon/Villeurbanne, France), partnership between CNRS/IN2P3 and CEA/DSM/Irfu, and computing resources supported by the Chinese Academy of Sciences.


\bibliographystyle{elsarticle-num} 


\appendix


\section{Coordinate transformations}
\label{appnd1}

\renewcommand{\theequation}{A\arabic{equation}}
\renewcommand{\thefigure}{A\arabic{figure}}
\renewcommand{\thetable}{A\arabic{table}}
\setcounter{figure}{0} 
\setcounter{table}{0} 

The ECEF system is a right-handed Cartesian coordinate system with its origin fixed at the center of the Earth, and rotates along with it. The x-axis passes through the equator at the Prime Meridian, the y-axis passes through the equator 90$^\circ$ east of the Prime Meridian, and the z-axis passes through the North Pole. The \texttt{ECEF} class takes instances of other known coordinate systems as arguments and transforms them into ECEF system. All properties of the \texttt{CartesianRepresentation} class are inherited by \texttt{ECEF}.

The geodetic coordinate system is defined in the \texttt{GeodeticRepresentation} class and has \texttt{latitude}, \texttt{longitude}, and \texttt{height} as attributes. Its origin is at the center of the Earth; the geodetic system uses the same coordinate frame as ECEF. Locations in the Northern and Southern Hemispheres have positive and negative \texttt{latitude} values, respectively. The \texttt{latitude} of the South Pole is -90$^\circ$, the North Pole is 90$^\circ$, and the equator is 0$^\circ$.  Values of \texttt{longitude} are positive towards East and negative towards West, ranging from 0$^\circ$ to 360$^\circ$. The \texttt{height} coordinate represents the height above the ellipsoidal Earth model or the mean sea level. Its value can range from \unit[-6378137]{m} to positive infinity. The ellipsoidal Earth model is a mathematical surface defined by a semi-major axis and a semi-minor axis that corresponds to the mean sea level. The most common values for these two parameters, used in {\sc GRANDlib}, are defined by the World Geodetic Standard 1984 (WGS-84)~\cite{10.1007/BFb0011360}. In {\sc GRANDlib}, the geodetic coordinate system is built using the \texttt{GeodeticRepresentation} class.

The local tangential plane (LTP) is a right-handed Cartesian coordinate system with a user-defined origin and orientation. Its coordinate frame is fixed with respect to the Earth and rotates along with it. The \texttt{LTP} class takes instances of other coordinate systems as arguments and transforms them into the LTP system.
Commonly used orientations for the LTP frame are North-West-Up (NWU) and East-North-Up (ENU). Any combination of East (E), West (W), North (N), South (S), up (U), and down (D) in right-handed Cartesian is a valid orientation. The \texttt{LTP} class inherits all features of the \texttt{CartesianRepresentation} class. GRANDCS is a subclass of LTP (Section~\ref{sec:coordinate}).

{\sc GRANDlib} transforms coordinates between coordinate representations---i.e., between Cartesian and spherical representations---and between coordinate systems. Reference~\cite{481290} describes coordinate transformation from the ECEF to geodetic systems. A coordinate in the geodetic system with longitude $\phi$, latitude $\theta$, and height $h$ is transformed into $x$, $y$, and $z$ coordinates in the ECEF system as
\begin{align}
    x &= (r + h) \, {\rm cos}(\theta) \, {\rm cos}(\phi) \;, \notag \\
    y &= (r + h) \, {\rm cos}(\theta) \, {\rm sin}(\phi) \;, \notag \\
    z &= [r\,(1 - e ^ 2) + h] \, {\rm sin}(\theta) \;, \notag
\end{align}
where $r \equiv a  [1-(e^2 \sin^2 \theta)]^{-1}$, $a = \unit[6378137]{m}$ is the semi-major axis of the WGS-84 ellipsoid, and $e = 0.081819190842622$ is the eccentricity of the WGS-84 ellipsoid.

{\sc GRANDlib} internally performs coordinate transformations from the LTP and GRANDCS systems to other coordinate systems using the transformation matrix constructed from the basis vectors defined in the ECEF frame. The basis vectors depend on the location and orientation of the LTP or GRANDCS frame. For example, the basis vectors of the ENU orientation of the LTP of GRANDCS frame at a location with longitude $\phi$ and latitude $\theta$ are given in the ECEF frame as
\begin{align*}
    \boldsymbol{E} =
    \begin{bmatrix}
        -{\rm sin}(\phi) \\
        {\rm cos}(\phi) \\
        0 \\
    \end{bmatrix},\
    \boldsymbol{N} =
    \begin{bmatrix}
        -{\rm sin}(\theta)\,{\rm cos}(\phi) \\
        -{\rm sin}(\theta)\,{\rm sin}(\phi) \\
        {\rm cos}(\theta)\\
    \end{bmatrix},\
    \boldsymbol{U} =
    \begin{bmatrix}
        {\rm cos}(\theta)\,{\rm cos}(\phi) \\
        {\rm cos}(\theta)\,{\rm sin}(\phi) \\
        {\rm sin}(\theta) \\
    \end{bmatrix} \;.
\end{align*}
The corresponding transformation matrix is
\begin{equation}
    \boldsymbol{R} =
    \begin{bmatrix}
        E_x & E_y & E_z\\
        N_x & N_y & N_z\\
        U_x & U_y & U_z\\
    \end{bmatrix} \;.
    \label{mat:basis}
\end{equation}
A coordinate vector expressed in the ECEF frame, $\boldsymbol{V}_{\rm ECEF}$, is transformed into one expressed in the ENU frame, $\boldsymbol{V}_{\rm ENU}$, via
\begin{equation}
    \boldsymbol{V}_{\rm ENU}
    =
    \boldsymbol{R}
    \cdot
    \boldsymbol{V}_{\rm ECEF} \;,
    \label{eq:venu}
\end{equation}
and, conversely,
\begin{equation}
    \boldsymbol{V}_{\rm ECEF}
    =
    \boldsymbol{R}^{\rm T}
    \cdot
    \boldsymbol{V}_{\rm ENU} \;.
    \label{eq:vecef}
\end{equation}

The transformation matrix, $\boldsymbol{R}$, also allows us to compute the basis vectors of a coordinate frame with any orientation at a given location. For example, in order to compute the basis vectors of an LTP or GRANDCS system with NWU orientation, we define the vectors $\boldsymbol{N}$, $\boldsymbol{W}$, and $\boldsymbol{U}$ in the frame with ENU orientation.  They are then transformed into the ECEF frame using Eq.~(\ref{eq:vecef}), after which they serve as the basis vectors for the NWU orientation and are used to construct its transformation matrix. With it, transforming coordinates between that frame and the ECEF frame is performed similarly to the process described for the ENU orientation in Eqs.~(\ref{eq:venu}) and (\ref{eq:vecef}).

The transformation from LTP and GRANDCS systems to the geodetic system is performed in two steps. First, coordinates in LTP and GRANDCS are transformed into ECEF, following the procedure described above.  Then the coordinates in ECEF are transformed into geodetic. Similarly, coordinates in geodetic are transformed into LTP and GRANDCS frame in two steps using ECEF as an intermediate step. The same technique is used to transform coordinates from one LTP frame into another.


\section{Data format}
\label{appnd:fileformat}

\renewcommand{\theequation}{B\arabic{equation}}
\renewcommand{\thefigure}{B\arabic{figure}}
\renewcommand{\thetable}{B\arabic{table}}
\setcounter{figure}{0} 
\setcounter{table}{0} 

Data from GRAND DUs and GRAND simulated data are stored in ROOT \texttt{TTrees}. The {\sc GRANDlib} Data-Oriented Interface (DOI) module can read and write these data. This module allows users to work with GRAND data using only Python, without the user needing to work directly with ROOT classes. All the data can be read and written as standard Python scalars, lists, and arrays; the DOI internally converts them to the C++ variable types required by the \texttt{TTrees}. The DOI is contained in the \texttt{root\_trees} module; it is described in detail in Ref.~\cite{Piotrowski:2023xbc}.

\begin{table}[]
    \centering
    \begin{tabular*}{0.485\textwidth}{l p{6.3cm} l }
        \hline
        Class name & \hspace{2.45cm} Function \\
        \hline
        \hline
        \texttt{TRun} & Run information common to all events \\
        \texttt{TRunVoltage} & Voltage information common to all events of a run\\
        \texttt{TADC} & Traces of an event in ADC units before conversion to voltage\\
        \texttt{TRawVoltage} & Traces of an event after conversion to voltage\\
        \texttt{TVoltage} & Voltage traces of an event at the antenna feed point\\
        \texttt{TEfield} & Electric-field traces of an event\\
        \texttt{TShower} & Characteristics of a shower\\
        \texttt{TShowerSim} & Additional information about a simulated shower\\
        \texttt{TRunNoise} & Information about noise common to all events of a run\\
        \hline
    \end{tabular*}
    \caption{\textbf{Classes in the {\sc GRANDlib} \texttt{root\_trees} module and their functions.}  Different classes are used to store and access data related to extensive air showers, simulated or detected.}
    \label{tab:class_name}
\end{table}

Table~\ref{tab:class_name} shows the major data classes contained in the \texttt{root\_trees} module.
An ``event'' is usually a candidate for an extensive air shower generated by a cosmic ray, gamma ray, or neutrino. A ``run'' is defined as a collection of events detected within a given time span.  \texttt{TTrees} named with the \texttt{TRun*} pattern, where \texttt{*} is an alphanumeric wildcard, contain information that remains constant for the duration of the run, each entry in \texttt{TTrees} corresponding to a separate run.  In contrast, each entry in \texttt{TTrees} named with the \texttt{T*} pattern (without ``\texttt{Run}'') corresponds to a separate event. The classes \texttt{T*ADC}, \texttt{T*RawVoltage}, \texttt{T*Voltage}, \texttt{T*Efield}, and \texttt{T*Shower} contain information about ADC counts, the voltage measured at the antenna footpoint, the voltage at the antenna arms, electric fields at the antenna arms, parameters of the extensive air shower, respectively. The class \texttt{T*Sim} contains information exclusively generated by simulators. The class \texttt{T*} (without ``\texttt{Sim}'') contains information coming from hardware (which can also come from simulation, leaving unavailable fields empty).


\section{Code quality}
\label{appnd4}

\renewcommand{\theequation}{C\arabic{equation}}
\renewcommand{\thefigure}{C\arabic{figure}}
\renewcommand{\thetable}{C\arabic{table}}
\setcounter{figure}{0} 
\setcounter{table}{0} 

{\sc GRANDlib} includes tools to assess and maintain code quality. They are based on widely used packages like \texttt{pylint}, \texttt{coverage.py}, \texttt{mypy}, and \texttt{SonarQube}.

The \texttt{pylint} package is used as a static code analysis tool to check if the code meets PEP8 standards and detect errors. It reduces repetition by spotting code duplication. To improve coding standards, {\sc GRANDlib} also uses the \texttt{mypy} package for type annotation. It is a static type checker for Python.

The \texttt{coverage.py} package measures the code coverage of each module in {\sc GRANDlib}, i.e., the ratio of the amount of executed \textit{vs.}~non-executed code in a test script.
The coverage is low for a module whose test script does not cover the entire code.
At the time of writing, the overall code coverage of {\sc GRANDlib} is 84\%.
During the debugging phase, it is possible to run tests only for the module that the user is working on.


\section{{\sc GRANDlib} package dependencies}
\label{appnd5}

\renewcommand{\theequation}{D\arabic{equation}}
\renewcommand{\thefigure}{D\arabic{figure}}
\renewcommand{\thetable}{D\arabic{table}}
\setcounter{figure}{0} 
\setcounter{table}{0} 

Table~\ref{tab:pckg_depnd} lists the external software packages that {\sc GRANDlib} depends on.  They are automatically fetched and installed as part of the {\sc GRANDlib} installation~\cite{grandlib_repository}.  The $\nu_\tau$ interaction package {\sc DANTON} and the air-shower simulation packages {\sc CoREAS} or {\sc ZHAireS} must be installed and run separately.  See the documentation in Ref.~\cite{grandlib_repository}.

\begin{table}[t!]
    \centering
    \begin{tabular*}{\columnwidth}{l l l l }
        \hline
        black    & jupyter-notebook & paramiko  & SonarQube \\
        cffi     & lxml             & psycopg2  & sshtunnel \\
        coverage & matplotlib       & pylint    & TURTLE \\
        docker   & mypy             & ROOT      & typing\_extensions \\
        GULL     & numpy            &  &  \\
        \hline
    \end{tabular*}
    \caption{\textbf{External packages used by {\sc GRANDlib}.} These packages are automatically included in the provided docker environment wherein {\sc GRANDlib} is run. See Ref.~\cite{grandlib_repository} for details.}
    \label{tab:pckg_depnd}
\end{table}


\end{document}